\documentclass[12pt]{iopart}
\usepackage{iopams,amsthm} 
\newtheorem*{lem}{Lemma}
\newtheorem*{thm}{Theorem}
\newtheorem*{prop}{Proposition}
\begin{document}

\title[The robustness of a many-body decoherence formula of Kay]
{The robustness of a many-body decoherence formula of Kay under changes 
in graininess and shape of the bodies}

\author{Varqa Abyaneh and Bernard S Kay}

\address{Department of Mathematics, University of York, York YO10 5DD, UK}

\ead{pha98va@yahoo.co.uk and bsk2@york.ac.uk}

\begin{abstract}
In ``Decoherence of macroscopic closed systems within Newtonian quantum
gravity'' (Kay B S 1998 {\it Class. Quantum Grav.} {\bf 15} L89-L98) it
was argued that, given a many-body Schr\"odinger wave function $\psi(
{\bf x}_1, \dots, {\bf x}_N)$  for the centre-of-mass degrees of freedom
of a closed system of $N$ identical uniform-mass balls of mass $M$ and
radius $R$, taking account of quantum gravitational effects and then
tracing over the gravitational field amounts to multiplying the
position-space density matrix   $\rho({\bf x}_1, \dots, {\bf x}_N;  {\bf
x}_1', \dots, {\bf x}_N')= \psi({\bf x}_1, \dots, {\bf x}_N)\psi^*({\bf
x}_1', \dots, {\bf x}_N')$  by a multiplicative factor, which, if the
positions $\lbrace {\bf x}_1, \dots , {\bf x}_N; {\bf x}_1',
\dots , {\bf x}_N'\rbrace$ are all much further away from one another
than $R$, is well-approximated by\hfil\break $\left (\prod_K\left(|{\bf
x}_K-{\bf x}_K'|\over R\right)\prod_{I<J} \left({|{\bf x}_I'-{\bf
x}_J||{\bf x}_I-{\bf x}_J'|}\over {|{\bf x}_I-{\bf x}_J||{\bf x}_I'-{\bf
x}_J'|}\right)  \right)^{-24M^2}$. Here we show that if each
uniform-mass ball is replaced by a grainy ball or more general-shaped
lump of similar size consisting of a number, $n$, of well-spaced small
balls of mass $m$ and radius $r$ and, in the above formula, $R$ is
replaced by $r$, $M$ by $m$ and the products are taken over all $Nn$
positions of all the small balls, then the result is well-approximated
by replacing $R$ in the original formula by a new value
$R_{\hbox{eff}}$.  This suggests that the original formula will apply
in general to physically  realistic lumps -- be they macroscopic lumps
of ordinary matter with the grains atomic nuclei etc. or be they atomic
nuclei themselves with their own (quantum) grainy substructure -- provided $R$
is chosen suitably.   In the case of a cubical lump consisting of
$n=(2\ell+1)^3$ small balls ($\ell\ge 1$) of radius $r$ with centres at
the vertices of a cubic lattice of spacing $a$ (assumed to be very much
bigger than $2r$) and side $2\ell a$ we establish the bound $e^{-1/3}
(r/a)^{1/n}\ell a \le R_{\hbox{eff}} \le 2\sqrt 3 (r/a)^{1/n} \ell a$.
\end{abstract}


\maketitle

\section{Introduction and Results}

In \cite{kay1}, one of us argued that several of the puzzles inherent
in our current understanding of quantum and quantum-gravitational
physics would appear to find a natural resolution if one were to
postulate that quantum gravity is a quantum theory of a conventional
type with a total Hilbert space $H_{\hbox{total}}$ which arises as a
tensor product
\[H_{\hbox{total}}=H_{\hbox{matter}}\otimes H_{\hbox{gravity}}\]
of a matter and a gravity Hilbert space,  and with
a total time-evolution which is unitary, but that, while, as would
usually be assumed in a standard quantum theory, one still assumes
that there is an ever-pure time-evolving ``underlying'' state, modelled
by a density operator of form
\[\rho_{\hbox{total}}=|\Psi\rangle\langle\Psi|,\]
$\Psi\in H_{\hbox{total}}$, at each ``instant of time'',  one adds the
new assumption that the {\it physically relevant} density operator is
not this underlying density operator, but rather its partial trace,
$\rho_{\hbox{matter}}$, over $H_{\hbox{gravity}}$.   With these
assumptions, an initial underlying state of a closed  quantum
gravitational system with a low degree of matter-gravity entanglement
would be expected to become more and  more entangled as time increases
-- equivalently $\rho_{\hbox{matter}}$ will be subject to an
ever-increasing amount of decoherence.  In fact the von-Neumann entropy
of $\rho_{\hbox{matter}}$ would be expected to increase monotonically,
thus (when the theory is applied to a model for the universe as a whole)
offering an explanation for the Second Law of Thermodynamics and (when
the theory is applied to a model closed system consisting of a black
hole sitting in an otherwise empty universe) offering a resolution to
the Information Loss Puzzle.  (It offers both of these things in that,
by defining the physical entropy of $\rho_{\hbox{total}}$ to be the
von-Neumann entropy of $\rho_{\hbox{matter}}$, one reconciles an
underlying unitary time-evolution on $H_{\hbox{total}}$ with an entropy
which varies/increases in time.)

A second paper, \cite{kay2}, by the same one of us investigated the implications
of the theory proposed in \cite{kay1} for the decoherence of ordinary matter in
the non-relativistic, weak-gravitational-field regime.  In particular,
\cite{kay2} addressed the question of the appropriate description of the state,
at some instant of time, of the center-of-mass degrees of freedom of a
system of $N$ bodies, which, for simplicity, were taken to be identical.
In ordinary quantum mechanics, such a state would be described by a
many-body Schr\"odinger wave function $\psi({\bf x^1},\dots, {\bf x^N})$
in the usual matter Hilbert space $H_{\hbox{matter}}$, consisting of the
(appropriately symmetrized, according to whether the bodies are treated
as fermions or bosons or neither) $N$-fold tensor product of $L^2(R^3)$.
\cite{kay2} argues that such a $\psi$ needs to be replaced by a total
(entangled) matter-gravity state vector $|\Psi\rangle$ in a total 
matter-gravity Hilbert space which is the tensor product of  
$H_{\hbox{matter}}$ (defined as above) with a suitable Hilbert space 
$H_{\hbox{gravity}}$ for the modes of the quantized linearized gravitational
field.  Specifically, thinking of this tensor-product Hilbert space as the set
of (appropriately symmetrized) square-integrable functions from 
$R^{3N}$ to $H_{\hbox{gravity}}$, $|\Psi\rangle$ is taken, in \cite{kay2},
to be the function
\[({\bf x}_1, \dots, {\bf x}_N) \mapsto
\psi({\bf x^1}, \dots {\bf x^N}) |\gamma({\bf x^1}, \dots {\bf
x^N})\rangle\]
where $|\gamma({\bf x^1}, \dots {\bf x^N})\rangle$ is a certain
(non-radiative) quantum coherent state (introduced in \cite{kay2}) of
the linearised gravitational field describing the Newtonian
gravitational field due to the simultaneous presence of a body centred
on each of the locations ${\bf x^1}, \dots {\bf x^N}$.  (We remark that,
in \cite{kay2}, for the purposes of calculating these coherent states,
the bodies are modelled as classical mass-distributions.) In the
non-relativistic, weak-gravitational-field regime, the projector
$\rho_{\hbox{total}}=|\Psi\rangle\langle\Psi|$ onto this $|\Psi\rangle$,
is thus taken to be, to a very good approximation, the correct
description of the ``underlying'' state of quantum gravity describing
our many-body system in the sense described in the previous paragraph.
The physically relevant density operator is therefore given, in this
approximation, by the partial trace $\rho_{\hbox{matter}}$ of  this
$\rho_{\hbox{total}}$ over $H_{\hbox{gravity}}$ which, in position space
is clearly given by
\[ \fl \rho_{\hbox{matter}}({\bf x}_1, \dots , {\bf x}_N; {\bf x}_1', \dots ,
{\bf x}_N') \]
\[ = \psi({\bf x}_1, \dots , {\bf x}_N)\psi^*({\bf x}_1', \dots , {\bf x}_N')
{\mathcal M}({\bf x}_1, \dots , {\bf x}_N; {\bf x}_1', \dots , {\bf x}_N')\]
where the {\it multiplicative factor}
${\mathcal M}({\bf x}_1, \dots , {\bf x}_N; {\bf x}_1', \dots , {\bf
x}_N')$ is defined by
\[{\mathcal M}({\bf x}_1, \dots , {\bf x}_N; {\bf x}_1', \dots , {\bf
x}_N') = \langle\gamma({\bf x}_1, \dots , {\bf x}_N)|
\gamma({\bf x}_1', \dots , {\bf x}_N')\rangle.\]
(In \cite{kay2}, $\mathcal M$ was written $e^{-D}$ where $D$ was called
the {\it decoherence exponent}.) Moreover, it was shown in \cite{kay2}
that, in the case the bodies are taken to be balls with constant mass
density and (we work in Planck units where $G=c=\hbar=1$) total mass $M$
and radius $R$  (i.e. in the case the classical mass distributions
representing the bodies are taken to be such balls) and in the case all
the positions $\lbrace {\bf x}_1,\dots , {\bf x}_N; {\bf x}_1', \dots ,
{\bf x}_N'\rbrace$ of their centres of mass are much further away from
one another than $R$ (we shall call this the {\it well-spaced}
regime)\footnote{In many applications, one expects the region of
$(\hbox{configuration space})\times(\hbox{configuration space})$ where
this condition doesn't hold to be so small in comparison to size of the
region where $\psi\psi^*$ is significantly big that no significant
correction would be needed to results calculated on the assumption that
the multiplicative factor is well-approximated by the ${\mathcal M}_a$
in all of $(\hbox{configuration space})\times(\hbox{configuration
space})$.}, the multiplicative factor is well approximated by ${\mathcal
M}_a$ where ${\mathcal M}_a$ is given by the explicit formula
\begin{equation}
\label{eqn:1.1}
{\mathcal M}_a({\bf x}_1, \dots , {\bf x}_N; {\bf x}_1', \dots , {\bf x}_N')=
\prod_{I=1}^N\prod_{J=1}^N\left ({|{\bf x}_I'-{\bf
x}_J||{\bf x}_I-{\bf x}_J'|\over |{\bf x}_I-{\bf x}_J||{\bf x}_I'-{\bf
x}_J'|}\right )^{-12M^2}
\end{equation}
where it is to be understood that, in the cases $I=J$, the terms in the
denominator
\break $|{\bf x}_I-{\bf x}_J||{\bf x}_I'-{\bf x}_J'|$ are to be replaced
by $R^2$.

To summarize, taking into account gravitational effects and then tracing
over the gravitational field, as the general theory of \cite{kay1}
prescribes that we should do, has, in the non-relativistic
weak-gravitational-field regime, and on the assumption that the bodies
are uniform mass balls, an overall effect which is equivalent to
multiplying the usual quantum mechanical (pure) position-space density
matrix $\psi({\bf x}_1, \dots , {\bf x}_N) \psi^*({\bf x}_1', \dots ,
{\bf x}_N')$ by a multiplicative factor, $\mathcal M$, which, in the
well-spaced regime, is well-approximated by ${\mathcal M}_a$ given by
the formula (\ref{eqn:1.1})  and it is the product of this pure density
matrix with the multiplicative factor which is to be regarded as the
physically relevant density operator.

We remark that the formula (\ref{eqn:1.1}) can also be written
\[ \fl {\mathcal M}_a({\bf x}_1, \dots , {\bf x}_N; {\bf x}_1', \dots ,
{\bf x}_N')= \left (\prod_K\left(|{\bf x}_K-{\bf x}_K'|\over
R\right)\prod_{I<J} \left({|{\bf x}_I'-{\bf x}_J||{\bf x}_I-{\bf
x}_J'|}\over {|{\bf x}_I-{\bf x}_J||{\bf x}_I'-{\bf x}_J'|}\right) 
\right)^{-24M^2}\]
where the first product is taken over all $K$ from $1$ to $N$ and the
second product is taken over all $I$ and $J$ from 1
to N, which satisfy $I<J$.
We also remark that, in the case of a wave function for a single ball state,
this prescription amounts to multiplying the density operator
$\psi({\bf x})\psi({\bf x}')$ by the multiplicative factor
\[{\mathcal M}_a({\bf x}; {\bf x'})= (|{\bf x}-{\bf x}'|/R)^{-24M^2}.\]

The problems we wish to address in the present paper are that, in the
derivation, \cite{kay2}, of (\ref{eqn:1.1}), the bodies are assumed to
be constant mass-density balls whereas in applications, the bodies one
is interested in will actually typically be ``grainy'', and also not
necessarily spherical.  In fact, in one application (cf. the discussion
of the Schr\"odinger-cat-like states in \cite{kay2}) of the above
formulae, the balls are interpreted as macroscopic balls of ordinary
matter and real ordinary matter is of course grainy in that it is made
out of atoms etc. and we may also be interested in lumps of ordinary
matter with shapes other than balls. In another application, the above
formulae (or rather their obvious generalization to states of many
bodies where the masses and radii are not all equal) are interpreted as
telling us how the standard non-relativistic quantum mechanics of closed
systems of large numbers of atomic nuclei and electrons etc. gets
modified according to the theory developed in \cite{kay1} and
\cite{kay2}.   In this latter application, the balls are taken to be
models of nuclei and electrons etc. and again, of course, real nuclei
and electrons are not actually uniform density balls, but will have
their own grainy substructure and also need not be spherical.  

However one might hope that if constant-density balls were replaced by
grainy, and possibly non-spherical, (classical) ``lumps'', then the
formula $(\ref{eqn:1.1})$ for the multiplicative factor might
nevertheless remain approximately correct provided we replace $R$ in
$(\ref{eqn:1.1})$ by a suitable effective radius $R_{\hbox{eff}}$.  One
might further hope that, in the case of a grainy ball,  $R_{\hbox{eff}}$
would turn out to be of the same order of magnitude as $R$ and in the
case of a grainy lump of some other shape, to be of the same order of
magnitude as some measure of the lump's typical linear size.  (We shall
refer below to the lump's {\it diameter} without attempting to give a
precise general definition to this notion.) If these hopes were
fulfilled, then one might say that the formula $(\ref{eqn:1.1})$ for
${\mathcal M}_a$ is {\it robust} under (classical) changes in graininess
and in shape.  

We remark that we are continuing to assume here that our grainy,
non-spherical, situations can, as far as their gravitational effects are
concerned, still be modelled within the formalism of \cite{kay2}, as
classical (albeit now no longer uniform and no longer spherical) mass
distributions and we admit that, in principle, we should of course
presumably work within an extension of the formalism of \cite{kay2}
which allows for truly {\it quantum} (and also relativistic)
descriptions of graininess (and non-sphericity). (Especially, such a
quantum relativistic description may well be important in the second of
our applications, to the grainy substructure of the proton etc.)  While
we shall not attempt to explore such quantum notions of graininess  and
non-sphericity in this paper, if robustness holds in the above sense
under classical changes in graininess and in shape, it might also seem
reasonable to guess that our formulae will remain robust even if
graininess and changes in shape were taken into account in such a
correct quantum way.  

In any case, the purpose of the present paper is to investigate whether
and/or under what circumstances, these hopes are fulfilled in the case
of one specific class of classical grainy models. Namely, models in
which every one of the balls of mass $M$ and radius $R$ in formula
$(\ref{eqn:1.1})$ is replaced by a collection of $n$ little balls, each
of mass $m$ and radius $r$ fixed at definite positions inside a
lump-shaped region (i.e. some smooth closed bounded region of ${\bf
R}^3$) of diameter approximately equal to $2R$ so that its centre of
mass is located where the center (i.e. one of the positions ${\bf x}_1,
\dots, {\bf x}_N$) of the ball it replaces was located. We shall assume
that the configurations of the balls within each of the lumps are
Euclidean-congruent to one-another. However we shall not restrict the
relative orientations of the different lumps; they are allowed to be
arbitrarily  rotated with respect to one another.

We remark that obviously we have $nm=M$. Here we have in mind choices of
values for the various parameters such that \begin{equation}
\label{eqn:1.2} \fl r \quad \hbox{$\ll$ any inter-small-ball spacing}
\quad and \quad R \quad\hbox{$\ll$ any inter-lump spacing},
\end{equation} where, by ``any inter-lump spacing'' we mean any of the
distances between pairs of distinct elements of the set $\lbrace {\bf
x}_1, \dots , {\bf x}_N; {\bf x}_1', \dots , {\bf x}_N'\rbrace$ and
(assuming from now on that a  definite system has been adopted for
enumerating the little balls  -- as $x_{I1}, \dots, x_{In}$ etc. --
within the $I$th lump etc.) by ``any inter-small-ball spacing'' we mean
any of the distances between pairs of distinct elements of the set
$\lbrace {\bf x}_{I1}, \dots , {\bf x}_{In}; {\bf x}_{I1}', \dots , {\bf
x}_{In}'\rbrace$ for some/any $I$. Then (in all but an unimportantly
small volume of $(\hbox{configuration space})\times(\hbox{configuration
space})$ -- cf. footnote above) formula $(\ref{eqn:1.1})$ will clearly
get replaced by 
\begin{equation} \label{eqn:1.3} \fl {\mathcal
M}_a^{\hbox{grainy}}({\bf x}_1, \dots , {\bf x}_N; {\bf x}_1', \dots ,
{\bf x}_N')= \prod_{I=1}^N\prod_{i=1}^n\prod_{J=1}^N\prod_{j=1}^n\left
({|{\bf x}_{Ii}'-{\bf x}_{Jj}||{\bf x}_{Ii}-{\bf x}_{Jj}'|\over |{\bf
x}_{Ii}-{\bf x}_{Jj}| | {\bf x}_{Ii}'-{\bf x}_{Jj}'|}\right )^{-12m^2}
\end{equation} 
where now it is to be understood that in the cases where
$I=J$ {\it and} $i=j$, the terms in the denominator $|{\bf x}_{Ii}-{\bf
x}_{Jj}||{\bf x}_{Ii}'-{\bf x}_{Jj}'|$ are to be replaced by $r^2$.

In line with $(\ref{eqn:1.2})$, if we make the replacements:
\[{\bf x}_{Ii}'-{\bf x}_{Jj}\rightarrow {\bf x}_{I}'-{\bf x}_{J}
,\quad\hbox{and}\quad {\bf x}_{Ii}-{\bf x}_{Jj}\rightarrow {\bf x}_{I}-{\bf
x}_{J} \quad\hbox{{\it except} when $I=J$}\]
(and similarly with primed and unprimed quantities interchanged) in
$(\ref{eqn:1.3})$, then we clearly expect to get a good
approximation\footnote
{In fact this will clearly be true in the sense that if we scale the
positions of the centres of mass of the lumps while not scaling the
lumps themselves, then  ${\mathcal M}^{\hbox{grainy}}_a/{\mathcal M}_a$
tends to $1$ as the scale tends to infinity.} to ${\mathcal
M}_a^{\hbox{grainy}}$. Making these replacements, we immediately get
\begin{eqnarray}
\fl
{\mathcal M}_a^{\hbox{grainy}}({\bf x}_1, \dots , {\bf x}_N; {\bf x}_1', \dots ,
{\bf x}_N') \nonumber\\
\lo{\simeq} \left(\left({\prod_I\prod_J|{\bf x}_I'-{\bf x}_J|^{n^2}
| {\bf x}_I-{\bf x}_J'|^{n^2}}\over{\prod_{I\ne J} |{\bf x}_I-{\bf
x}_J|^{n^2}|{\bf x}_I'-{\bf x}_J'|^{n^2}}\right )\prod_{i,j}
\left ({1\over|{\bf x}_i-{\bf x}_j|}\right )^{2N} \right )^{-12m^2} \nonumber\\
\lo= \left(\left({\prod_I\prod_J|{\bf x}_I'-{\bf x}_J|
| {\bf x}_I-{\bf x}_J'|\over\prod_{I\ne J} |{\bf x}_I-{\bf
x}_J||{\bf x}_I'-{\bf x}_J'|}\right )\prod_{i,j}
\left ({1\over|{\bf x}_i-{\bf x}_j|}\right )^{2N/n^2} \right )^{-12M^2}
\nonumber
\end{eqnarray}
where the final product involves all the positions ${\bf x}_i$ of little
balls inside any single lump -- by our assumption of Euclidean
congruence, it doesn't matter which one -- and ranges over all values of
$i$ and $j$ from $1$ to $n$ {\it except that} it is to be understood
that, in this final product, in the cases $i=j$ the denominator is to be
replaced by $r^2$.  Clearly another way of saying exactly the same thing
as this is that ${\mathcal M}_a^{\hbox{grainy}}$ is approximately equal
to ${\mathcal M}_a$, as given by the formula $(\ref{eqn:1.1})$ {\it
except that} the proviso that, in the cases $I=J$, the terms in the
denominator $|{\bf x}_I-{\bf x}_J||{\bf x}_I'-{\bf x}_J'|$ are to be
replaced by $R^2$ should be {\it replaced by} the proviso that, in the
cases $I=J$, the terms in the denominator
$|{\bf x}_I-{\bf x}_J||{\bf x}_I'-{\bf x}_J'|$ are to be replaced by
$R_{\hbox{eff}}^2$ where $R_{\hbox{eff}}$ is defined by
\[R_{\hbox{eff}}=\left (\prod_i\prod_j|{\bf x}_i-{\bf x}_j|\right )^{1\over
n^2}\]
where the products each go from $1$ to $n$ and it is to be understood
that, in cases where $i=j$, the terms $|{\bf x}_i-{\bf x}_j|$ are to be
replaced by $r$.  Equivalently, we may write
\begin{equation}
\label{eqn:1.4}
R_{\hbox{eff}}=\left (r^n\prod_{i<j}|{\bf x}_i-{\bf x}_j|^2\right
)^{1\over n^2}
\end{equation}
(where the product is now over all $i$ and $j$ from $1$ to $n$
satisfying $i < j$). Thus the first part of our hope
(i.e. that replacing our uniform density balls by our grainy lumps
can be well-approximated by replacing $R$ by a suitable $R_{\hbox{eff}}$)
is fulfilled.

We now turn to discussing  the second part of our hope, namely whether,
and/or under what circumstances, $R_{\hbox{eff}}$ will turn out  to be
of the same ``order of magnitude'' as the diameter of our lump.  We have
not been able to answer this question in general but, instead, content
ourselves with analysing one particularly simple special case of our
model.  Namely, where each lump is a cube consisting of $n=(2\ell+1)^3$
small balls of radius $r$ with centres at the vertices of a cubic
lattice of spacing $a$  and side $2\ell a$. To spell out precisely what
we mean and at the same time set up a useful notation, we assume the
ball-centres of any one of these cubes to be coordinatizable so that
they lie at the positions $(i_1a,i_2a,i_3a)$ where $i_1$, $i_2$, and
$i_3$ are integers which each range between $1$ and $2\ell+1$.  Clearly
the total number of balls, $n$, in the cube will be $(2\ell + 1)^3$ and
we note that there will be a ball at the centre (with coordinates
$((\ell +1)a, (\ell + 1)a, (\ell + 1)a)$).   In line with
$(\ref{eqn:1.2})$, we assume
\[r \ll a \quad \hbox{and} \quad R
\quad\hbox{$\ll$ any inter-lump spacing}.\]
(Of course, we require $r < a/2$ for the balls to fit into the lattice
at all!)

For such a lump, $(\ref{eqn:1.4})$ can clearly be rewritten in the form
\begin{equation}
\label{eqn:1.5}
R_{\hbox{eff}}=\left (r\over a \right )^{1/n}\left (\prod_{i<j}\left |
{{\bf x}_i-{\bf x}_j}\over a \right |^2\right )^{1\over n^2}a
\end{equation}
(here we continue to assume some system has been adopted for numbering
our balls from $1$ to $n=(2\ell +1)^3$) and we notice that, for many
values of the pair $(r/a, n)$, the prefactor, $(r/a)^{1/n}$ will itself
be of order 1. This will be the case if, as is relevant to a ball or
lump of ``ordinary matter'', we take  $r/a$ to be of the order of the
ratio of the radius of the proton to the Bohr radius (i.e. around
$10^{-5}$) even for $\ell=1$ ($n=27$) and in fact, if $\ell$ is $2$
($n=125$) or more (and it will be much more if we are taking our cube to
be a model for a macroscopic lump of ordinary matter as considered in
\cite{kay2}) then this prefactor will in fact be a number close in value
to 1. On the assumption that this prefactor is of order 1 (or very close
to 1) our question then reduces to the question whether the quantity
\[\left (\prod_{i<j}\left | {{\bf x}_i-{\bf x}_j}\over a \right
|^2\right )^{1\over n^2}\] is of the same order as $\ell$.  Choosing
units such that $a=1$, this amounts to a question about the
quantity\footnote{It is necessary to bear in mind, here
and elsewhere, that $n$ is shorthand for $(2\ell+1)^3$.}
\begin{equation}
\label{eqn:1.51}
Q(\ell)=\left (\prod_{i<j}\left |
{\bf x}_i-{\bf x}_j \right |\right )^{2\over n^2}
\end{equation}
where the product is over all pairs of distinct points of the cubic
lattice of points with integer coordinates $(i_1,i_2,i_3)$ where $i_1$,
$i_2$ and $i_3$ range between $1$ and $2\ell+1$ (from now on we shall
often simply call such a finite cubic lattice of points a ``cube'') and,
again, we assume that some system for numbering these points from $1$ to
$n=(2\ell+1)^3$ has been adopted. We formulate this question in a
precise way by asking whether one can find numbers, $c_1$, $c_2$, of
order 1 such that
\begin{equation}
\label{eqn:1.6}
c_1 \ell \le Q(\ell) \le c_2 \ell.
\end{equation}
It is easy to see that one can satisfy the upper bound by choosing
$c_2=2\sqrt 3$.  To see this, notice that each term in the product in
$(\ref{eqn:1.51})$ is obviously less than or equal to $2\sqrt 3\ell$ since
this is the length of the body-diagonal of the cubic lattice and
moreover there are $n(n-1)/2$ terms in the product.  We thus
have that $Q(\ell) \le (2\sqrt 3 \ell)^{n(n-1)/n^2}$ which, since $\ell \ge
1$, is clearly less than $2\sqrt 3 \ell$.

However, as far as we can see, to show that there exists a (non-zero)
positive $c_1$ such that $c_1\ell$ is a lower bound is not completely
trivial because $Q(\ell)$ is the $2/n^2$ power (i.e. $2/(2\ell+1)^3$
power) of a product of numbers which range in magnitude from 1 to
numbers of order $\ell$ (the largest of the numbers will of course be
$2\sqrt 3 \ell$).  Nor, to our knowledge, does it easily follow from any
existing standard mathematical result\footnote{As far as we are aware,
the notion of closest relevance to this question in the mathematical
literature is the notion \cite{Goluzin} of the {\it transfinite
diameter} of a closed bounded region  (which one shows, see again
\cite{Goluzin}, is equivalent to another notion known as the  {\it
capacity} of that region) of the (complex) plane.  This is defined to be
the supremum over all finite sets of points in the region of the
$1/n(n-1)$ power of the product of all the distances between pairs of
distinct points of the set. (One can show, for example, that the
transfinite diameter of a disk is equal to its radius.)  If we allow
ourselves to generalize this concept of transfinite diameter to closed
bounded regions of {\it three}-dimensional Euclidean space, then our
$Q(\ell)$ may be seen to resemble one of the terms in the supremum which
would define the transfinite diameter of a cube of side $2\ell + 1$
(except that the $1/n^2$ power is taken, instead of the $1/n(n-1)$
power).  However, our interest is in $Q(\ell)$ itself and no supremum is
to be taken.} However, we have succeeded in finding the (albeit probably
 not best-possible) number $c_1 = e^{-1/3}$ for which we can prove that
it does hold:

\begin{prop}
\[
e^{-1/3} \ell \le Q(\ell) \le  2\sqrt 3\ell.
\]
\end{prop}
\noindent
(Actually, one can show \cite{Abyaneh}, by similar methods to those in
the proof of the lower bound below, that if we omit $\ell=1$ the upper
bound of $2\sqrt 3 \ell$ can be reduced to $2.61 \ell$.)  We present our
proof of the (lower bound in) the proposition in a separate section below.
Putting this proposition together
with $(\ref{eqn:1.5})$, we thus have  the following theorem.

\begin{thm}

For a lump consisting of $n=(2\ell +1)^3$ small balls of mass $m$ and radius
$r$, centred at the vertices of a cubical region (as specified above) of
side $2\ell a$ of a cubic lattice of spacing $a$, 
\[e^{-1/3} (r/a)^{1/n}\ell a
\le R_{\hbox{eff}} \le 2\sqrt 3 (r/a)^{1/n} \ell a.\] 
(Here, we recall that
$a$ is of course assumed to be greater than $2r$ and $R_{\hbox{eff}}$,
when inserted in place of $R$ in $(\ref{eqn:1.1})$, is expected to give
a good approximation to the multiplicative factor $M_a^{\hbox{grainy}}$
of $(\ref{eqn:1.3})$ if $a \gg 2r$.)

\end{thm}

We remark that it is easy to calculate $Q(\ell)$ numerically for small
values of $\ell$ (say for $\ell=1\dots 10$) and the numerical evidence
suggests that,  for such small values of $\ell$, $Q(\ell)$ is 
well-approximated by a formula of form $Q(\ell)\simeq A\ell + B$, i.e.
\begin{equation}
\label{eqn:1.7}
Q(\ell)/\ell \simeq A+B/\ell
\end{equation}
where $A \simeq 1.2$ and $B \simeq 0.6$.  If it could be proven that there
exist exact values for $A$ and $B$ such that this approximate formula holds for
all $\ell$ with an error term which tends to zero as $\ell$ tends to infinity,
then this would of course be a stronger result than our proposition 
above and would, in particular, tell us that our value for $c_1$, 
$e^{-1/3}\simeq 0.71$, can be improved to $1.2$.  
However, we have been unable to prove this.

\section{Proof of Proposition}

We have already proven the upper bound above, so it remains to prove the
lower bound. For
this proof, we find it useful to write $Q(\ell)$ as the product
\[Q(\ell)=\prod_{i=1}^n\omega_i(n)\]
where
\[\omega_i(n)=\prod_{j\ne i}\left (\left |
{\bf x}_i-{\bf x}_j \right |\right )^{1\over n^2}\]
where the product is over all $j$ from $1$ to $n$ except that the factor
with $j=i$ is omitted. We shall also assume that the numbering of the
vertices from $i=1$ to $n$ is such that the centre vertex (with
coordinates, as introduced above, $(\ell, \ell, \ell)$) is numbered
$i=1$.

\begin{lem}

\[ \omega_1 \le \omega_j \quad (j=1, \dots, n).\]

\end{lem}

\begin{proof} In terms of our coordinatization of our cube, we need to
show that the product

\[\mu_{(j_1,j_2,j_3)}=\prod_{(i_1,i_2,i_3)\ne
(j_1,j_2,j_3)}|(j_1,j_2,j_3)-(i_1,i_2,i_3)|\]
over all distances from an arbitrary lattice point $(j_1,j_2,j_3)$ to all
the other lattice points in our cube is greater than or equal to the product
\[\mu_{(\ell+1,\ell+1,\ell+1)}=\prod_{(i_1,i_2,i_3)\ne (\ell+1,\ell+1,\ell+1)}
| (\ell+1,\ell+1,\ell+1)-(i_1,i_2,i_3)|\]
over all distances from the centre point, with coordinates
$(\ell+1,\ell+1,\ell+1)$, to all the other lattice points in our cube. 
(To explain the notation here,
$\omega_j=(\mu_{(j_1,j_2,j_3)})^{1/n^2}$ if $(j_1,j_2,j_3)$ are the
coordinates of the point numbered $j$.  The inequality of the lemma will
then follow by taking the $1/n^2$ power of each side of the above inequality.)
We will prove this by exhibiting, for each $(j_1,j_2,j_3)$, a one-one
onto mapping $f_{(j_1,j_2,j_3)}$ from (coordinatized) lattice points of
our cube to themselves with the properties
\begin{equation}
\label{eqn:2.1}
f_{(j_1,j_2,j_3)}((j_1,j_2,j_3))=(\ell+1,\ell+1,\ell+1)
                        \end{equation}
and
			\begin{equation}
\label{eqn:2.2}
|f_{(j_1,j_2,j_3)}((i_1,i_2,i_3))-(\ell+1,\ell+1,\ell+1)|
\leq |(i_1,i_2,i_3)-(j_1,j_2,j_3)|
			\end{equation}
In fact, given such a mapping we will immediately have
\begin{eqnarray}
\fl \mu_{(j_1,j_2,j_3)}=\prod_{(i_1,i_2,i_3)\ne (j_1,j_2,j_3)}
| (j_1,j_2,j_3)-(i_1,i_2,i_3)| \nonumber\\
\lo{\ge} \prod_{(i_1,i_2,i_3)\ne (j_1,j_2,j_3)}
| f_{(j_1,j_2,j_3)}((j_1,j_2,j_3))-f_{(j_1,j_2,j_3)}((i_1,i_2,i_3))|
\nonumber\\
\lo= \prod_{(i_1,i_2,i_3)\ne (j_1,j_2,j_3)}
| (\ell+1,\ell+1,\ell+1)-f_{(j_1,j_2,j_3)}((i_1,i_2,i_3))| \nonumber\\
\lo= \prod_{(\hat i_1,\hat i_2, \hat i_3)\ne (\ell+1,\ell+1,\ell+1)}
| (\ell+1,\ell+1,\ell+1)-(\hat i_1,\hat i_2, \hat i_3)| \nonumber\\
\lo= \mu_{(\ell+1,\ell+1,\ell+1)} \nonumber\\
\end{eqnarray}
where, in the penultimate equality, we have defined $(\hat i_1,\hat i_2, \hat
i_3)$ to be $f_{(j_1,j_2,j_3)}((i_1,i_2,i_3))$.

It therefore remains to exhibit a mapping, $f_{(j_1,j_2,j_3)}$, with the above
properties.  To do this, we define
\[ \fl f_{(j_1,j_2,j_3)}((i_1,i_2,i_3))=
(\ell+1-j_1+i_1,\ell+1-j_2+i_2,\ell+1-j_3+i_3)
 \ (\hbox{mod} \ 2\ell+1)\]
where we use a non-conventional notion of {\it addition modulo $q$} in which
$q \ (\hbox{mod} \ q)$ is deemed to be $q$ rather than $0$. One may
understand this definition geometrically as a rigid translation of all
the points of our cube where, however, if the would-be destination of a
point is outside of our cube, the point instead gets mapped to the
counterpart-position in our cubic lattice to the position it would
arrive at in a neighbouring cube, were our cube to be surrounded by
similar neighbours in a cubic lattice arrangement.

Clearly this is a bijection and satisfies (\ref{eqn:2.1}).  To show it
satisfies (\ref{eqn:2.2}) we calculate, using the obvious properties of
our notion of ``mod'':
\begin{eqnarray}
\fl
|f_{(j_1,j_2,j_3)}((i_1,i_2,i_3))-(\ell+1,\ell+1,\ell+1)|
= ( \left[( (\ell+1-j_1 +i_1) (\hbox{mod}  \ 2\ell+1))
-\ell-1 \right]^2 \nonumber\\
\lo+ \left[ ((\ell+1-j_2 +i_2) (\hbox{mod}  \ 2\ell+1))
-\ell-1 \right]^2 \nonumber\\
\lo+ \left[( (\ell+1-j_3 +i_3) (\hbox{mod} \ 2\ell+1))
-\ell-1 \right]^2 )^{1/2} \nonumber\\
\lo{\leq} \left( (j_1-i_1)^2+(j_2-i_2)^2+(j_3-i_3)^2 \right)^{1/2}
= |(j_1,j_2,j_3)-(i_1,i_2,i_3)|. \nonumber
\end{eqnarray}
\end{proof}

Before proceeding with the proofs of the two bounds in our proposition, 
it will be useful to define, for a given cube (i.e. cubic lattice of
points) of side $2\ell + 1$, certain special sets of lattice points. 
First, we define the ``$k$th centre shell'' ($k\leq \ell$) to be the set
of lattice points on the surface of the cube of side $2k+1$ centred on
the centre of our given cube (so that a cube of side $2\ell+1$ would
have a total of $\ell$ centre shells).  

\begin{prop} [Lower Bound] \label{prop:lower}

For a cube of side $2\ell+1$, we have
\begin{equation}
\label{eqn:cor}
\frac{Q(\ell)}{\ell} \geq e^{-\frac{1}{3}} \ \ \ \ \forall \ell \in \mathbb{N}
\end{equation}

\end{prop}

\vspace{0.25cm}

\begin{proof}

For such a cube we have that
\[
n= (2\ell+1)^3
\]
Let us denote the number of lattice points in the $k$th centre shell by $n_k$.
This is given by
\[
n_k= 6(2k+1)^2-12(2k+1)+8
\]
We first use the lemma  to estimate
\[
Q(\ell) = \prod_{i=1}^{n} \omega_i(n) \geq \omega_1(n)^{n}
\]
We can rewrite $\omega_1 (n)$ by a relabelling of the lattice points
\begin{equation}
\label{eqn:rhoi}
\omega_1(n)=\left( \prod_{i=1}^{n} |{\bf x}_1-{\bf x}_i|
\right) ^{\frac{1}{n^2}}= \left( \prod_{k=1}^\ell \prod_{d=1}^{n_k} 
|{\bf x}_1-{\bf x}_{d_k}|  \right)^{\frac{1}{n^2}}
\end{equation}
where ${\bf x}_{d_k}$ is the coordinate-triple of the $d^{th}$ lattice
point in the $k$th centre shell. Each lattice point in the $k$th centre
shell is at least a distance $k$ away from the centre lattice point so
we have
\begin{equation}
\label{eqn:suff}
Q(\ell) \geq  \alpha (\ell)^{\frac{1}{n}}
\end{equation}
where
\begin{equation}
\label{eqn:alpha}
\alpha(\ell) = \prod_{k=1}^\ell k^{n_k}
\end{equation}
From (\ref{eqn:suff}), it can be seen that to prove (\ref{eqn:cor}) it
will be sufficient to prove
\begin{equation}
\label{eqn:prop}
\alpha (\ell) \geq \left(e^{-\frac{1}{3}}\ell \right)^{n}=
\left(e^{-\frac{1}{3}}\ell \right)^{(2\ell+1)^3}
\end{equation}
This can be proved by induction on $\ell$.  For all $\ell \in
\mathbb{N}$, let $P(\ell)$ be the proposition that (\ref{eqn:prop}) is
true. We have
\[
\alpha(1)=1 \geq e^{-\frac{1}{3}},
\]
so $P(1)$ is true.  So, if we can show $P(\ell) \Rightarrow P(\ell+1)$
($\forall \ell \in \mathbb{N}$) then the proposition will have been
proved to be true. From (\ref{eqn:alpha}) and (\ref{eqn:prop})
\[
\alpha(\ell+1)=\alpha(\ell) (\ell+1)^{n_{\ell+1}} \geq \left(
e^{-\frac{1}{3}}\ell \right)^{n} (\ell+1)^{n_{\ell+1}}
\]
Thus to now prove our proposition it will be sufficient to prove that
\[
\left( e^{-\frac{1}{3}}\ell
\right)^{n} (\ell+1)^{n_{\ell+1}} \geq \left( e^{-\frac{1}{3}} 
(\ell+1)  \right)^{(2\ell+3)^3}
\]
This is easily seen to be equivalent to
\[
e^{-\frac{1}{3}} \leq \left( 1+\frac{1}{\ell}  \right)^{-\frac{1}{3}\ell+
\left( \frac{132\ell^2+72\ell-1}{24\ell^2} \right)}.
\]
However the above equation clearly holds so we have proved our inductive
proposition (and hence the lower bound in our main proposition).
 
\end{proof}

\ack
We thank an anonymous referee for pointing out to us that numerical evidence
suggests the linear form for $Q(\ell)$ of equation $(\ref{eqn:1.7})$.
VA is grateful to Richard Hunter for useful discussions about, and
constructive criticism of, some of the mathematical proofs in this
paper.  VA is also grateful to Jim Brink for advice on numerical computation.
VA also thanks PPARC for a research studentship.  BSK is
grateful to Richard Hall for telling him about the concept of {\it
transfinite diameter} ($=$ {\it capacity}) of regions of the plane.  BSK
would also like to thank the Leverhulme foundation for a Leverhulme
fellowship (RF\&G/9/RFG/2002/0377) from October 2002 to June 2003 during
the course of which this research was begun.

\end{document}